\documentclass[usenatbib,useAMS,fleqn]{mn2e}

\usepackage{epsf}
\usepackage{amssymb}
\usepackage{amsmath}

\newcommand{\diff}{\ensuremath{{\rm d}}}

\newcommand{\Lstar}{\ensuremath{L_{\ast}}}
\newcommand{\Rstar}{\ensuremath{R_{\ast}}}
\newcommand{\Mstar}{\ensuremath{M_{\ast}}}

\newcommand{\Teff}{\ensuremath{T_{\rm eff}}}

\newcommand{\rfreq}{\ensuremath{\Omega}}
\newcommand{\rfreqc}{\ensuremath{\Omega}_{\rm c}}
\newcommand{\drfreq}{\ensuremath{\bar{\Omega}}}

\newcommand{\veq}{\ensuremath{V_{\rm e}}}
\newcommand{\vcrit}{\ensuremath{V_{\rm c}}}

\newcommand{\pfreq}{\ensuremath{\sigma}}

\newcommand{\dfreq}{\ensuremath{\omega}}
\newcommand{\dfreqr}{\ensuremath{\omega_{\rm r}}}
\newcommand{\dfreqi}{\ensuremath{\omega_{\rm i}}}
\newcommand{\dfreqz}{\ensuremath{\omega_{0}}}

\newcommand{\pperi}{\ensuremath{P_{\rm i}}}

\newcommand{\dmfreq}{\ensuremath{\delta\omega}}

\newcommand{\growth}{\ensuremath{\eta}}

\newcommand{\elle}{\ensuremath{\tilde{\ell}}}

\newcommand{\disc}{\ensuremath{D}}

\newcommand{\Yml}{\ensuremath{Y^{m}_{\ell}}}
\newcommand{\lamml}{\ensuremath{\lambda^{m}_{\ell}}}

\newcommand{\nrad}{\ensuremath{n}}

\newcommand{\work}{\ensuremath{W}}
\newcommand{\dwork}{\ensuremath{\diff W/\diff x}}
\newcommand{\Epul}{\ensuremath{E_{W}}}

\newcommand{\Msun}{\ensuremath{{\rm M}_{\sun}}}

\newcommand{\days}{\ensuremath{\rm d}}
\newcommand{\hours}{\ensuremath{\rm h}}

\newcommand{\poinc}{Poincar\'{e}}

\newcommand{\boojum}{\textsc{boojum}}

\title[Kappa-mechanism excitation of retrograde mixed modes]
      {Kappa-mechanism excitation of retrograde mixed modes 
       in rotating B-type stars}
\author[R.H.D.Townsend]
       {R. H. D. Townsend$^{1,2}$%
        \thanks{Email: rhdt@bartol.udel.edu}\\
        $^{1}$ Bartol Research Institute,
        University of Delaware,
        Newark, DE 19716, USA\\
        $^{2}$ Department of Physics \& Astronomy, 
        University College London, 
        Gower Street, London WC1E 6BT}
\date{%
Received: .................................... 
Accepted: ....................................
}

\pagerange{\pageref{firstpage}--\pageref{lastpage}}
\pubyear{2005}

\begin{document}


\maketitle

\label{firstpage}

\begin{abstract}
I examine the stability of retrograde mixed modes in rotating B-type
stars. These modes can be regarded as a hybrid between the Rossby
modes that arise from conservation of vorticity, and the \poinc\ modes
that are gravity waves modified by the Coriolis force. Using a
non-adiabatic pulsation code based around the traditional
approximation, I find that the modes are unstable in mid- to late-B
type stars, due to the same iron-bump opacity mechanism usually
associated with SPB and $\beta$ Cep stars. At one half of the critical
rotation rate, the instability for $m=1\ldots4$ modes spans the
spectral types B4 to A0. Inertial-frame periods of the unstable modes
range from 100 days down to a fraction of a day, while normalized
growth rates can reach in excess of $10^{-5}$.

I discuss the relevance of these findings to SPB and pulsating Be
stars, and to the putative Maia class of variable star. I also outline
some of the questions raised by this discovery of a wholly-new class
of pulsational instability in early-type stars.
\end{abstract}



\section{Introduction} \label{sec:intro}

In a recent paper \citep[][hereafter T05]{Tow2005}, I investigated the
influence that the rotation-originated Coriolis force exerts over the
self-excited oscillations of slowly pulsating B-type (SPB) stars. To
maintain a link with historical theoretical studies of SPB stars in
which the effects of rotation were overlooked
\citep[e.g.,][]{GauSai1993,Dzi1993}, the analysis was restricted to
those modes that transform continuously into ordinary g modes as the
rotation rate is reduced toward zero. Following the classification
scheme for low-frequency waves in rotating fluids
\citep[e.g.,][]{Gil1982}, these comprise the prograde and retrograde
\poinc\ modes, the prograde mixed modes, and the Kelvin modes.

The present paper extends the T05 study to one of the types of
pulsation mode found only in rotating stars: the retrograde mixed
modes, also referred to in the literature as mixed Rossby-\poinc\
modes \citep[e.g,][]{Lou2000}, mixed Rossby-gravity modes
\citep[e.g.,][]{Gil1982}, and Yanai modes
\citep[e.g.,][]{Tow2003a}. After providing a brief background to these
modes in Section~\ref{sec:background}, I use an approximation-based
method (Sec.~\ref{sec:method}), very similar to that developed in T05,
to investigate their stability in a range of B-type stellar models
(Sec.~\ref{sec:models}). The principal finding (Sec.~\ref{sec:calc})
is that retrograde mixed modes are unstable in mid to late B-type
stars, due to the same iron-bump $\kappa$ mechanism responsible for
the pulsation of SPB and $\beta$ Cep stars. I discuss and summarize
the significance of this result in Section~\ref{sec:discuss}.

\section{Background} \label{sec:background}

In order to understand the nature of retrograde mixed modes, and why
they might be susceptible to $\kappa$-mechanism excitation, it is
helpful to review their relationship to the differing classes of
low-frequency mode present in rotating stars. These classes comprise
the \poinc, Kelvin, and Rossby modes on the one hand, and the mixed
modes (both prograde and retrograde) on the other.

The \poinc\ modes rely on a combination of buoyancy and the Coriolis
force to restore displaced fluid elements to their equilibrium
position. They reduce to ordinary gravity modes in the non-rotating
limit, having an angular dependence described by the spherical
harmonics \Yml\ with harmonic degree $\ell$ and azimuthal order $m$
satisfying $-\ell+2 \leq m \leq \ell$. In light of this
correspondence, it is common to refer to the \poinc\ modes simply as g
modes \citep[e.g.,][]{LeeSai1987,Bil1996}.

However, such a designation groups the \poinc\ modes with the Kelvin
modes, since the latter \emph{also} reduce to g modes, having
$m=-\ell$, in the non-rotating limit. These Kelvin modes, which always
propagate in the prograde azimuthal direction\footnote{I adopt the
usual convention that negative $m$ corresponds to prograde
propagation, and positive $m$ to retrograde propagation, as viewed
from the co-rotating frame.}, have rather different properties than
the \poinc\ modes, making it desirable to maintain the distinction
between the two. Most significantly, Kelvin modes exhibit geostrophic
balance, where the polar component of the Coriolis force remains in
balance with the force arising from polar pressure gradients
\citep[see, e.g.,][]{Gil1982}. As a result, the horizontal propagation
of these modes is dispersion-free, distinguishing them from the
dispersive \poinc\ modes.

The Rossby modes \citep[or r modes; see][]{PapPri1978} are unlike the
\poinc\ and Kelvin modes, inasmuch as they do not require buoyancy to
restore displaced fluid elements. Instead, they rely on the
conservation of the radial component of total vorticity, operating on
fluid displacements across the curved level surfaces of the star's
stratification. [See \citet{Sai1982} for a particularly lucid
description of this process]. Rossby modes always propagate in the
retrograde direction, and in the limit of slow rotation follow the
first-order dispersion relation
\begin{equation} \label{eqn:r-freq}
\pfreq \approx \frac{2 m \rfreq}{\ell(\ell+1)}
\end{equation}
independent of the internal stellar structure. Here, and throughout
the present work, the symbols have the same meaning as in T05; in
particular, \pfreq\ is the pulsation angular frequency in the
co-rotating frame of reference, and \rfreq\ is the rotation angular
frequency. In the same slow-rotation limit, the horizontal velocity
fields generated by r modes approximate those produced by the trivial
toroidal modes found in non-rotating stars \citep[e.g.,][]{Unn1989},
with an angular dependence described by the curl of a radial vector
with length \Yml.

Mixed modes can be regarded as a hybrid between Rossby and \poinc\
modes. In the dispersion diagram for low-frequency modes in a rotating
star \citep[see][Fig.~1]{Lou2000}, they mark the dividing line between
the \poinc\ modes at higher \pfreq, and the Kelvin and Rossby modes at
lower \pfreq. At slow rotation rates, the prograde mixed modes behave
like \poinc\ modes, and are usually classified as the $\ell = -m+1$ g
modes. In contrast, the retrograde mixed modes have a Rossby-like
character, and are classified as r modes described by the dispersion
relation~(\ref{eqn:r-freq}) with $\ell = m$. As the rotation rate
increases, however, the prograde and retrograde mixed modes for each
value of $m$ approach one another in frequency, and exhibit properties
that lie somewhere between pure Rossby and \poinc\ modes.

There is a well-established body of literature on mixed modes in the
field of terrestrial atmospheric and oceanic geophysics, extending
back to their detection in the Earth's atmosphere by
\citet{YanMur1966}, coupled with the development by \citet{Mat1966} of
the `equatorial $\beta$-plane' approximation for analytical treatment
of low-frequency waves in rotating fluids. Literature discussing mixed
modes in a stellar astrophysical context is, however, comparatively
scarce. \citet{LeeSai1987,LeeSai1997} and \citet{DziKos1987} have both
noted the presence of r-mode solutions to the governing rotating-star
pulsation equations that acquire the character of g modes in the
limit of large spin parameter $\nu \equiv 2\rfreq/\pfreq$. Yet, a full
appreciation of the nature of these mixed-mode solutions did not
emerge until the $\beta$-plane approximation was applied to the
stellar pulsation problem, in the recent studies by \citet{Lou2000}
and \citet{Tow2003a}.

From a standpoint of stellar stability, the retrograde mixed modes are
particularly interesting on account of their relationship to r
modes. Being almost solenoidal \citep[e.g.,][]{Sai1982}, r modes
produce little compression or rarefaction of the stellar plasma, and
are therefore difficult to excite via the classical `Greek letter'
mechanisms \citep[$\gamma$, $\delta$, $\epsilon$ and $\kappa$;
see][]{Unn1989} which involve a thermodynamic Carnot cycle. This
explains why incidences of r-mode instability due to these mechanisms
are rare\footnote{It should be noted that there exist other,
non-classical mechanisms for the excitation of r modes \citep[e.g.,
driving by gravitational waves; see][]{And1998}, but these are
generally unimportant in the non-relativistic objects considered
here.}. Indeed, even in those situations where classical instability
has been found, the strength of the excitation -- as measured by its
linear growth rate -- has been much smaller than is typical for
ordinary p or g modes \citep[see, e.g.,][]{BerPro1983}.

However, in the case of retrograde mixed modes the non-solenoidal
\poinc\ character acquired with more-rapid rotation causes them to
generate appreciable compressions and rarefactions. If rotation is
sufficient, therefore, these modes stand a good chance of becoming
(significantly) unstable to one of the classical excitation
mechanisms. This recognition is the foundation for the present paper.

\section{Method} \label{sec:method}

The method used to investigate the stability of retrograde mixed modes
is based on the so-called `traditional approximation', and is very
similar to the approach described in detail in T05. In particular, I
solve an identical set of coupled differential equations~(7--10,
\emph{ibid}) and boundary conditions~(15--16, \emph{ibid}) as an
eigenproblem, using the non-adiabatic \boojum\ code.  The resulting
dimensionless eigenfrequencies
\begin{equation} \label{eqn:dfreq}
\dfreq \equiv \pfreq\,\sqrt{\frac{\Rstar^{3}}{G \Mstar}},
\end{equation}
define, via their imaginary parts \dfreqi, whether a mixed mode is
linearly stable ($\dfreqi > 0$) or unstable ($\dfreqi < 0$). The most
important departure from T05 relates to the calculation of the
effective harmonic degree \elle\ appearing in the pulsation
equations. In T05, \elle\ is obtained (eqn.~6, \emph{ibid}) from the
eigenvalue \lamml\ of Laplace's tidal equations that transforms
continuously into $\ell(\ell+1)$ in the limit of no rotation.

In the present analysis, I instead evaluate \elle\ from the family of
eigenvalues $\lambda$ that are appropriate to retrograde mixed
modes. Within the alternative indexing schemes introduced by
\citet{LeeSai1997} and \citet{Tow2003a}, they are labeled as the
$k=-1$ or $s=0$ solution branches of the tidal equations. Only minor
revisions to \boojum\ were required to permit consideration of these
branches, consisting primarily of alterations to the matrix routines
employed to calculate $\lambda$. However, initial usage of the updated
code revealed very poor convergence toward eigenfrequencies, with
thousands of relaxation iterations required to reach the desired
solution tolerance. After some investigation, the difficulty was
traced to the adoption in T05 of the steady-wave approximation. This
approximation, which entails using only the real part of the frequency
when evaluating $\nu$, permits easy solution of the tidal equations
since all variables remain real. However, it has the undesirable side
effect that the implicit dependence of $\lambda$ on \dfreq\ becomes
\emph{non-analytic} (i.e., $\lambda$ does not satisfy the
Cauchy-Riemann equations). Invariably, this tends to hamper the
convergence of \boojum, especially when considering mixed modes.

This difficulty is addressed within \boojum\ via a new two-stage
procedure for evaluation of $\lambda$. First, Laplace's tidal
equations are solved as before using only the real part \dfreqr\ of
the frequency. The solutions are then taken as the starting point in
an inverse iteration algorithm \citep[e.g.,][]{Pre1992} based on the
fully-complex frequency \dfreq. The algorithm is allowed to run until
the relative change in $\lambda$ reaches machine precision, which
typically takes five or six iterations. With this modification,
$\lambda$ behaves in an analytic fashion, and the convergence issues
are wholly resolved.

Further improvements to \boojum\ concern the discriminant root finding
process. Although the dimensionless eigenfrequencies of the retrograde
mixed modes remain defined by the roots
\begin{equation} \label{eqn:char}
\disc(\dfreq) = 0
\end{equation}
of the discriminant function (see T05, Sec.~2.3), I found that initial
isolation of roots was greatly facilitated by adopting the quantity
\begin{equation} \label{eqn:dm-freq}
\dmfreq \equiv \frac{\dfreqz - \dfreq}{\drfreq^{3}}
\end{equation}
as the independent variable, instead of \dfreq\ itself. This quantity,
also favoured by \citet{Sai1982}, furnishes a normalized measure of
departures from the first-order dispersion
relation~(\ref{eqn:r-freq}). The dimensionless rotation frequency
appearing in the denominator is defined as
\begin{equation}
\drfreq \equiv \rfreq\,\sqrt{\frac{\Rstar^{3}}{G \Mstar}},
\end{equation}
while the r-mode limit for mixed modes,
\begin{equation}
\dfreqz \equiv \frac{2\drfreq}{m + 1},
\end{equation}
comes from setting $\ell=m$ in the dispersion
relation~(\ref{eqn:r-freq}). Since deviations from sphericity arising
due to the centrifugal force are neglected in the present analysis,
\dfreq\ is always smaller than \dfreqz\ \citep[see, e.g.,][]{Sai1982},
and thus the `departure measure' \dmfreq\ is positive for all modes
considered.

The algorithm used by \boojum\ to solve the characteristic
equation~(\ref{eqn:char}) has also been altered. I found that the
complex secant approach favoured by \citet{Cas1971} consistently
outperforms the \citet{Mul1956} algorithm originally implemented, even
though the latter method has a higher (quadratic) order of convergence
toward roots.

\section{Stellar Models} \label{sec:models}

\begin{figure}
\epsffile{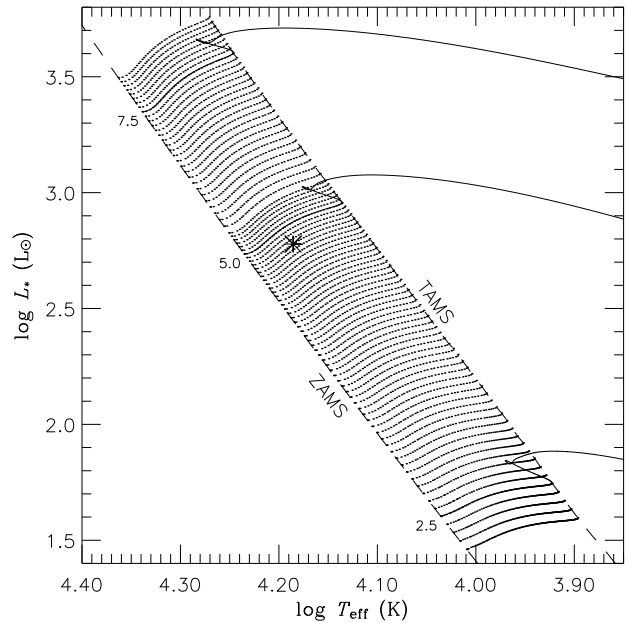}
\caption{The theoretical HR diagram for the stellar models introduced
in Section~\ref{sec:models}; these are plotted as points in the
effective temperature (\Teff) versus stellar luminosity (\Lstar)
plane. The dashed lines running diagonally from top left to bottom
right indicate the ZAMS and TAMS main sequence limits, while the
circle is centred on the position of the 53 Per model analyzed in
Section~\ref{ssec:53per}. Full evolutionary tracks for three selected
initial masses (labeled at the ZAMS in solar units) are also shown in
the diagram, drawn as solid lines.}
\label{fig:models}
\end{figure}

I use the Warsaw--New Jersey evolution code to calculate tracks of
stellar models for a metalicity $Z = 0.02$. These tracks, of which
there are 91 together comprising 4268 individual models, sample the
initial mass range $\Mstar = 2.3$--$5.3\,\Msun$ at a resolution
$0.05\,\Msun$ and the range $\Mstar = 5.3$--$8.3\,\Msun$ at a
resolution $0.1\,\Msun$; each extends from zero-age main sequence
(ZAMS) to terminal-age main sequence (TAMS). This coverage of the
theoretical Hertzsprung-Russel (HR) diagram, plotted in
Fig.~\ref{fig:models}, is shifted toward rather lower masses than
considered in T05, primarily to ensure -- based on initial exploratory
calculations -- that complete coverage of the mixed-mode instability
strips is secured (see Sec.~\ref{ssec:instab}). However, apart from
the differences in initial masses, the parameters and input data I
adopt for the stellar model calculations are identical to those
discussed in Section~3 of T05.

\section{Stability Calculations} \label{sec:calc}

\subsection{53 Per Model} \label{ssec:53per}

\begin{figure}
\leavevmode
\begin{center}
\epsffile{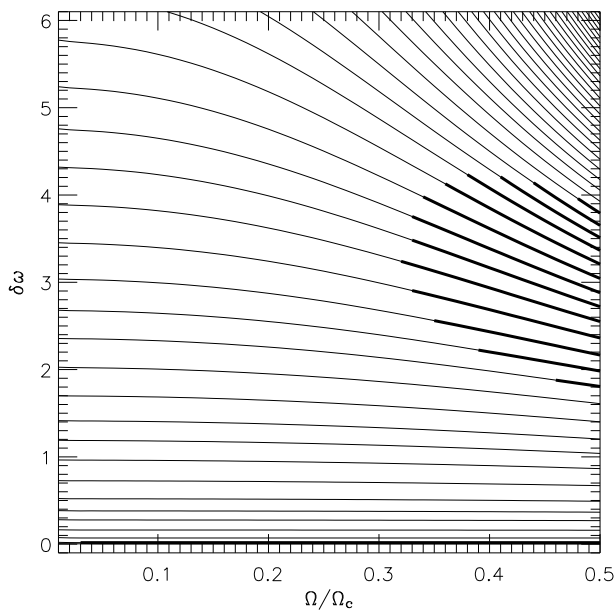}
\caption{Departure measures for $m=1$ mixed modes of the 53 Per
model, plotted as a function of rotation frequency. Each curve
corresponds to a particular radial order \nrad; the line weight is
used to indicate whether modes are stable (thin) or unstable (thick).}
\label{fig:53per}
\end{center}
\end{figure}

I first apply the updated \boojum\ code to investigate the stability
of $m=1$ mixed modes in the same 53 Per-like model introduced in
T05. A range of rotation angular frequencies, from the negligibly
rotating case up to the half-critical rate $\rfreq/\rfreqc=0.5$, is
considered. This upper bound is loosely dictated by the limitations of
the approximations described in Section~\ref{sec:method} (and see also
T05), but also happens to be close to the observed ceiling on SPB-star
rotation, $\rfreq/\rfreqc \approx 0.46$\footnote{In fact, this quoted
value should be taken as a lower estimate, since the inclination of
the star remains uncertain.}, established by the B5 pulsator HD~1976
\citep{Mat2001}.

The results from these stability calculations are plotted in
Fig.~\ref{fig:53per}, which traces the \rfreq\ dependence of the
eigenfrequency departure measure \dmfreq\ introduced in
eqn.~(\ref{eqn:dm-freq}), for modes of differing radial
orders\footnote{As enumerated by the number of nodes in the radial
displacement eigenfunction} \nrad. At slowest rotation, the flatness
of the \dmfreq\ curves indicates that the departure measure for each
mode is independent of \rfreq, confirming the finding by
\citet{PapPri1978} that deviations from the first-order dispersion
relation~(\ref{eqn:r-freq}) vary as $\mathcal{O}(\rfreq^{3})$. Toward
more-rapid rotation, higher-order terms become significant,
with the result that the curves shown in Fig.~\ref{fig:53per} dip
downward.

\begin{figure*}
\leavevmode
\begin{center}
\epsffile{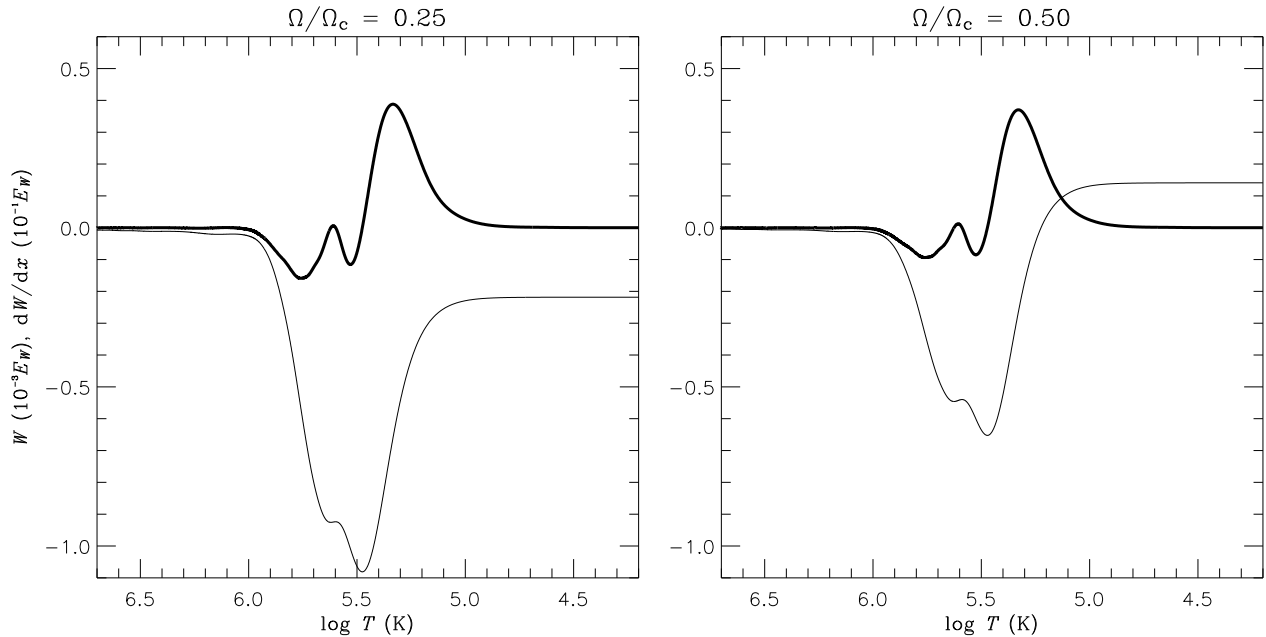}
\caption{The cumulative work (thin) and differential work (thick) for
the $m=1$, $\nrad=20$ mixed mode of the 53 Per model, plotted as a
function of temperature at rotation rates $\rfreq/\rfreqc = 0.25$
(left panel) and 0.5 (right panel). The work functions are expressed
in scaled units of the total energy of pulsation \Epul.}
\label{fig:work}
\end{center}
\end{figure*}

Of particular interest in the figure is the fact that a group of
modes, having radial orders $\nrad=13$--25, become unstable above a
rotation rate $\rfreq/\rfreqc \approx 0.3$. The nature of the
instability is revealed in Fig.~\ref{fig:work}, which plots both the
cumulative work \work\ and differential work \dwork\ for the
$\nrad=20$ mixed mode, at a rotation rate $\rfreq/\rfreqc = 0.25$ when
the mode is stable, and at the more-rapid rate $\rfreq/\rfreqc = 0.5$
when the mode has subsequently been destabilized. The sharp peak in
\dwork\ at $\log T \approx 5.3$ reveals that the mode is strongly
driven by the same iron-bump opacity mechanism considered responsible
for the p-mode instability of $\beta$ Cep stars
\citep{Cox1992,DziPam1993} and the g-mode instability of SPB stars
\citep{GauSai1993,Dzi1993}.

As elucidated by the latter-most authors \citep[and see
also][]{Pam1999}, two key requirements must be satisfied for the
$\kappa$ mechanism to be effective. First, the relative Lagrangian
pressure perturbation $\delta p/p$ should be large and vary slowly
with radius within the driving zone. This constraint on the shape of
the eigenfunction establishes a lower limit $\nrad \geq 13$ on the
radial order of unstable mixed modes in the 53 Per model, surprising
similar to the corresponding limit found in T05 for unstable g
modes. In fact, the coincidence between the two stems from the fact
that the same set of pulsation equations governs both classes of mode.

The second requirement is that the pulsation period should not be
significantly greater than the local thermal timescale within the
excitation zone. If instead the period is too long, radiative
diffusion between neighbouring fluid elements can lead to strong
damping in the inner envelope, that cancels out any driving due to the
$\kappa$ mechanism. This inhibition effect is illustrated in
Fig.~\ref{fig:work}. The strength of the $\kappa$-mechanism driving is
the same in both panels, as can be seen from the similar peaks $\diff
\work/\diff x \approx 0.04\,\Epul$ in the differential work at $\log T
\approx 5.3$. However, the radiative damping at $\log T \approx 5.7$
is more pronounced in the left-hand panel, owing to the longer period
of the mode -- $P=2.76\,\days$, as compared to $P=1.63\,\days$ for the
right-hand panel. This enhanced damping is sufficient to stabilize the
mode.

All mixed modes in Fig.~\ref{fig:53per} having $\rfreq/\rfreqc
\lesssim 0.3$ or $\dmfreq \gtrsim 4$ posses periods longer than the
thermal timescale in the driving zone. Hence, these modes are
stabilized by the same radiative damping described above. In the
following section, I explore how radiative damping, acting in tandem
with the restriction on the $\delta p/p$ eigenfunction morphology,
shapes which regions of the HR diagram exhibit mixed-mode instability
via the $\kappa$-mechanism. The remainder of the present section is
focused on a discussion of the unstable $\nrad=1$ mixed mode seen in
Fig.~\ref{fig:53per} as the thick line having $\dmfreq \approx 0$.

Inspection of the work function for this mode indicates that weak
driving in the core, at $\log T\approx 7.4$, is responsible for the
instability. Although this may be a result of genuine
$\epsilon$-mechanism excitation, the normalized growth rate of the
mode,
\begin{equation} \label{eqn:growth}
\growth \equiv -\dfreqi/\dfreqr,
\end{equation}
never exceeds $3.3\times10^{-13}$ at any rotation rate considered
(compare with the $\nrad=20$ mode, which reaches
$\growth=7.3\times10^{-6}$ at $\rfreq/\rfreqc=0.5$). This value
differs so little from zero that it would be more accurate to describe
the $\nrad=1$ mixed mode as neutrally stable. Such a stance is lent
support by the fact that the mode sits very close to the pure r-mode
limit $\dfreq = \dfreqz$, and -- being almost solenoidal -- can be
expected to be neither excited nor damped by classical mechanisms.

\subsection{Instability Strips} \label{ssec:instab}

\begin{figure*}
\leavevmode
\begin{center}
\epsffile{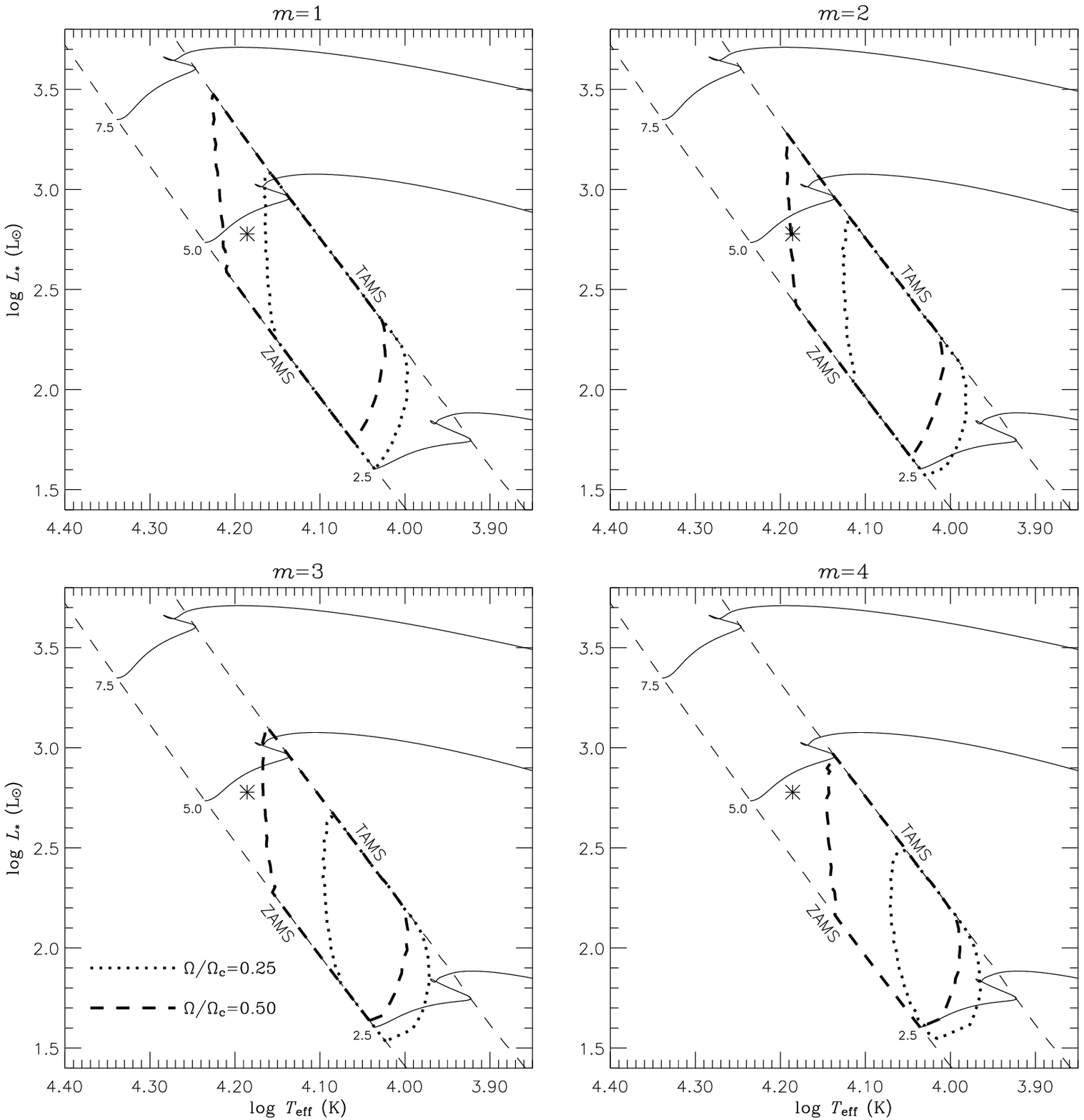}
\caption{Instability strips in the HR diagram, for $m=1\ldots4$ mixed
modes of the stellar models introduced in
Section~\ref{sec:models}. For each value of the azimuthal order, the
extent of the $\kappa$-mechanism instability at the rotation rates
$\rfreq/\rfreqc=0.25$ (dotted) and 0.5 (dashed) is indicated using
thick lines. The small open triangles, filled squares, and open
diamond show the inferred positions of selected SPB stars, pulsating
Be stars, and HD~208727, respectively (see text).}
\label{fig:instab}
\end{center}
\end{figure*}

\begin{figure*}
\leavevmode
\begin{center}
\epsffile{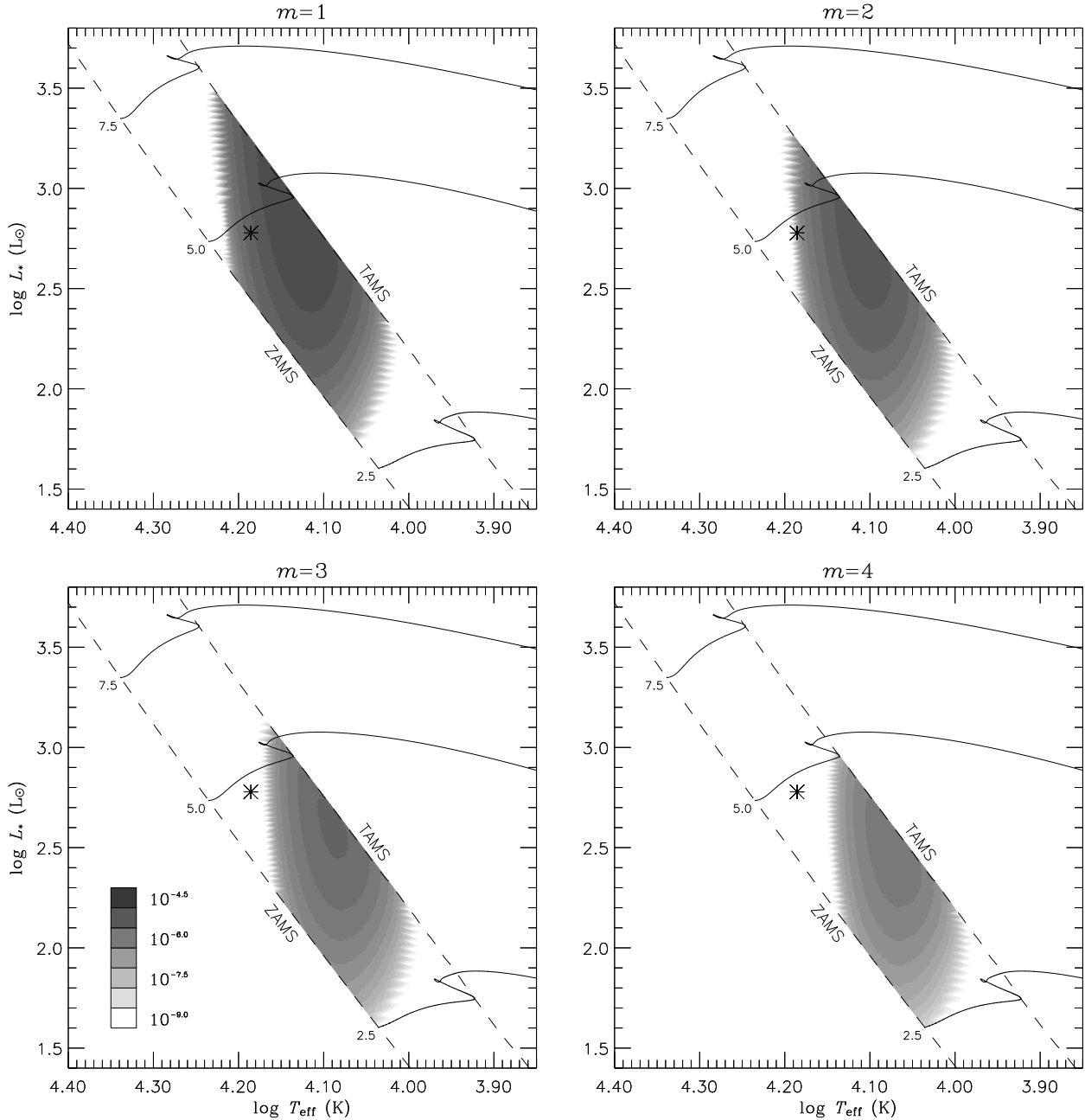}
\caption{Maximal growth rates in the HR diagram, for $m=1\ldots4$
mixed modes of the stellar models introduced in
Section~\ref{sec:models}. For each value of the azimuthal order, the
largest \growth\ at the rotation rate $\rfreq/\rfreqc=0.5$ is
indicated via the gray level. The distribution of these levels is
logarithmic, and extends from $10^{-9}$ (white) to $10^{-4.5}$
(black).}
\label{fig:growth}
\end{center}
\end{figure*}

\begin{figure*}
\leavevmode
\begin{center}
\epsffile{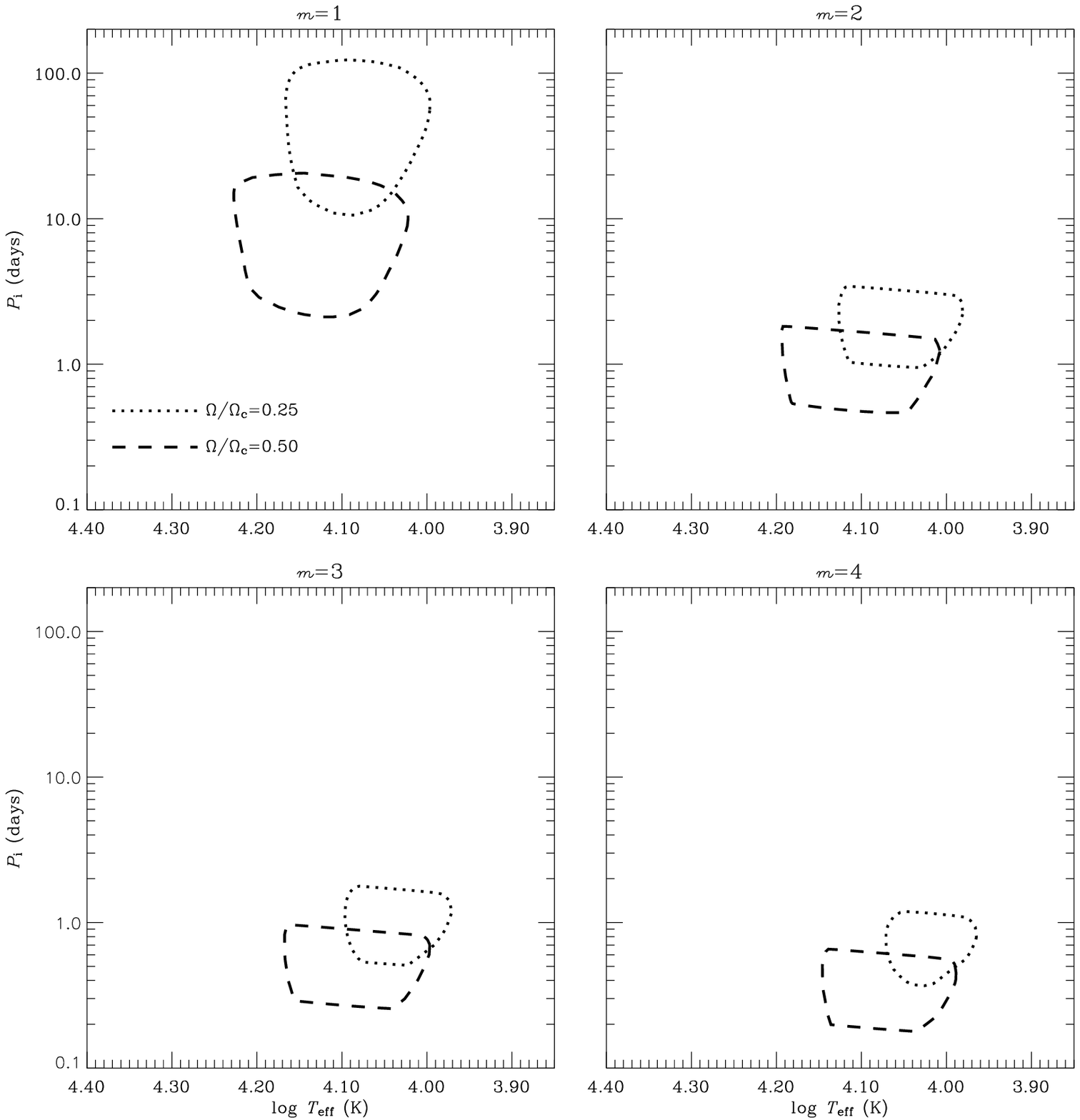}
\caption{Instability regions in the $\Teff-\pperi$ plane, for
$m=1\ldots4$ mixed modes of the stellar models introduced in
Section~\ref{sec:models}. For each value of the azimuthal order, the
extent of the $\kappa$-mechanism instability at the rotation rates
$\rfreq/\rfreqc=0.25$ (dotted) and 0.5 (dashed) is indicated using
thick lines. The small open triangles, filled squares, and open
diamond show the observed periods of selected SPB stars, pulsating Be
stars, and HD~208727, respectively (see text). Note that the ordinate
scale is logarithmic.}
\label{fig:period}
\end{center}
\end{figure*}

The scope of the investigation is now broadened, to encompass the
entire set of stellar models introduced in Section~\ref{sec:models}. I
use \boojum\ to investigate the stability of these models toward
$m=1\ldots4$ retrograde mixed modes. Fig.~\ref{fig:instab} plots the
instability strips in the HR diagram associated with these modes, at
two distinct rotation rates: $\rfreq/\rfreqc=0.5$ approximates the
observed upper limit on SPB-star rotation (see
Sec.~\ref{ssec:instab}), while $\rfreq/\rfreqc=0.25$ is representative
of more-moderately rotating stars. Formally, the condition of
instability is defined by the requirement that a mode has a positive
growth rate. However, it is appropriate to exclude those modes that,
similar to the $\nrad=1$ mode considered in the preceding section,
have such small values of \growth\ that they are more properly
regarded as neutrally stable. Therefore, the instability strips shown
in Fig.~\ref{fig:instab} are constructed to enclose all stellar models
that possess one or more modes with a growth rate in excess of the
threshold value $\growth=10^{-9}$.

Also shown in Fig.~\ref{fig:instab} are the inferred positions of
selected SPB and pulsating Be stars, and the position of the Maia
candidate star HD~208727, which is discussed further in
Section~\ref{sec:discuss}. The SPB data are compiled directly from the
tabulations by \citet{Wae1998}, \citet{Aer1999}, \citet{Mat2001} and
\citet{DeC2002}; in the many cases where multiple $(\Teff,\Lstar)$
values exist for a given star, the most recently published ones are
adopted. The pulsating Be-star data are obtained via a three-stage
process. First, the list of line-profile variable Be stars observed by
\citet{Riv2003} is merged with that of the short-period photometric
variables studied by \citet{Per2002,Per2004}. Then, the calibrations
by \citet{Cha2001} are used to determine the effective temperature and
surface gravity of each star in the merged list; if values are not
available for a given star, it is dropped from the list. Finally, I
define an empirical relation $\Lstar = \Lstar(\Teff,\log g)$ via a
quintic surface fit to the luminosity data of the stellar models
introduced in Section~\ref{sec:models}. This relation is used to map
the $(\Teff,\log g)$ values in the list into corresponding
$(\Teff,\Lstar)$ values, suitable for plotting in the HR diagram.

Figure~\ref{fig:instab} demonstrates that the mixed-mode instability
found in the 53 Per model (Sec.~\ref{ssec:53per}) extends to many
of the other stellar models considered. At a rotation rate
$\rfreq/\rfreqc=0.25$, the instability for the $m=1\ldots4$ modes
spans the effective temperature range $\log \Teff=3.96$--$4.17$. At
the more-rapid rate $\rfreq/\rfreqc=0.5$, the instability broadens and
shifts to toward higher effective temperatures and luminosities, with
a range $\log \Teff=3.99$--$4.23$ corresponding loosely to spectral
types B4 to A0 \citep{Boh1981}. At fixed rotation rate, incrementing
the azimuthal order displaces the instability in the opposite
direction, toward lower effective temperatures and luminosities.

The sensitivity of the instability toward both rotation rate and
azimuthal order comes from the period matching requirement discussed
in the preceding section. Returning once again to the first-order
dispersion relation~(\ref{eqn:r-freq}), it is clear that any increase
in \rfreq\ ($m$) will act to increase (decrease) the frequency of a
mixed mode, and hence to shorten (lengthen) its period. To ensure that
this period remains commensurate with the local thermal timescale in
the driving zone, the instability therefore shifts to hotter (cooler)
stars, in which the opacity bump is situated shallower (deeper) in the
envelope. A very similar line reasoning was employed in T05 to
understand the behaviour of the instability strips for \poinc, Kelvin
and prograde mixed modes.

\begin{table}
\leavevmode
\begin{center}
\caption{Proportions, expressed as percentages, of the SPB (Be) stars
that fall within the mixed-mode instability strips plotted in
Fig.~\ref{fig:instab}, for each azimuthal order and rotation rate
considered.} \label{tab:instab}
\begin{tabular}{@{}ccccc}
&
\multicolumn{4}{c}{$m$} \\
$\rfreq/\rfreqc$ &
1 & 2 & 3 & 4 \\ \hline
0.25 & 41 (3) & 13 (0) & 3 (0) & 1 (0) \\
0.50 & 67 (7) & 48 (3) & 41 (3) & 29 (3) \\
\end{tabular}
\end{center}
\end{table}

Examining the correspondence between the mixed-mode instability strips
and the positions of the selected pulsating stars plotted in
Fig.~\ref{fig:instab}, it is clear that there is not a great deal of
agreement between the two. This discrepancy is underlined by
Table~\ref{tab:instab}, which summarizes the proportion of stars
enclosed by the instability strips at each azimuthal order and
rotation rate considered. At best, the strips encompass only two
thirds of the SPB stars -- a reflection of the stars' bias toward the
late end of the B spectral type, whereas SPB stars tend to be
concentrated more around the mid-B types. In fact, the SPB stars
coincide far more closely with the theoretical instability strips predicted
for $\kappa$-mechanism excitation of ordinary g modes \citep[see][his
figs.~3 \&~4]{Pam1999}. Therefore, it seems unlikely that the SPB
phenomenon as a whole can be explained better by mixed modes than by
the widely-accepted g-mode interpretation.

The situation is superficially similar for the pulsating Be stars
shown in Fig.~\ref{fig:instab}; in no case do more than 7 percent of
these stars fall within the mixed-mode instability strips. In light of
the concentration of Be stars toward the early-B spectral types
\citep{Por1996}, and the conspicuous absence of any pulsating Be stars
later than spectral type B6 \citep[see][and references
therein]{Baa1989a,Baa1989b,PorRiv2003}, this result is hardly
surprising. However, in contrast to SPB stars, Be stars are known to
rotate at a significant fraction of their critical rate
\citep[$\langle \veq/\vcrit \rangle \approx 0.83$, corresponding to
$\langle \rfreq/\rfreqc \rangle \approx 0.96$;
see][]{Cha2001}. Although near-critical rotation lies beyond the reach
of the \boojum\ code, the general trends shown in
Fig.~\ref{fig:instab} suggest that as $\rfreq/\rfreq \rightarrow 1$
the mixed-mode instability strips should shift toward the early-B
spectral types, encompassing a much greater proportion of pulsating Be
stars. Therefore, it remains plausible that the variability of these
stars can be attributed to mixed modes, rather than the \poinc\ modes
that are usually supposed \citep[e.g.,][and references
therein]{Riv2003}. This possibility is returned to in
Section~\ref{sec:discuss}.

Having discussed the extent of the $\kappa$-mechanism instability, I
now shift the focus to its \emph{strength}. Fig.~\ref{fig:growth}
plots the maximal growth rates of the instability across the HR
diagram, for the $m=1\ldots4$ mixed modes at the rotation rate
$\rfreq/\rfreqc=0.5$. These values are intended to furnish some
measure of the robustness of the instability. To this end,
\citet{Dzi1993} employ the normalized growth rate introduced by
\citet{Ste1979}, but in the absence of any sound theory of non-linear
amplitude saturation, I prefer to adhere to the simple
definition~(\ref{eqn:growth}).

The general trend seen in the figure is that the growth rates are at
their greatest for the $m=1$ modes, reaching up to a value $\growth
\approx 10^{-4.5}$ that is commensurate with the strongest
instabilities associated with $\beta$ Cep and SPB stars. Toward larger
values of $m$, \growth\ declines rapidly, reflecting the fact that
mixed modes having larger azimuthal orders exhibit a more Rossby-like
character \citep[see, e.g.][Fig.~1; note that the horizontal
wavenumber $k_{x}$ plotted as abscissa is equivalent to $m$]{Lou2000},
and are therefore (Sec.~\ref{sec:background}) more difficult to excite
via the $\kappa$ mechanism.

A secondary trend in Fig.~\ref{fig:growth} is that the growth rates
appear systematically larger for the stellar models situated near the
TAMS. As shown in Fig.~4 of T05, these models are characterized by
driving-zone thermal timescales that are rather longer than the
corresponding ZAMS models. Presumably, mixed modes can satisfy the
period matching requirement more easily at these long timescales, and
therefore are less susceptible to inhibition arising from radiative
damping in the inner envelope.

To bring the present section to a close, Fig.~\ref{fig:period} plots
the mixed-mode instability regions in the $\Teff-\pperi$ plane. Here,
\begin{equation} \label{eqn:pperi}
\pperi \equiv -2\pi/(\pfreq - m\rfreq)
\end{equation}
is the pulsation period in an inertial frame of reference -- a
quantity of relevance to the observational community, since it can be
ascertained directly from time-series monitoring of variable
stars. The minus sign appears in this expression because, for the
mixed modes considered here, the co-rotating frame frequency \pfreq\
is always smaller than $m\rfreq$. Also shown in the figure are the
observed periods of the same stars plotted in Fig.~\ref{fig:instab};
in cases where a star is multiperiodic, only the principal period is
displayed.

The figure reveals that the inertial frame periods become shorter with
rotation rate, behaviour arising due to the appearance of \rfreq\ in
the denominator of the expression~(\ref{eqn:pperi}) for \pperi. More
significantly, it is evident that the unstable $m=1$ mixed modes
exhibit periods that are one or two orders of magnitude larger than
those with azimuthal orders $m \geq 2$. This is a consequence of the
fact that the $m=1$ modes have co-rotating frame frequencies that, to
first order in \rfreq, follow $\pfreq \approx \rfreq$ (see
eqn.~\ref{eqn:r-freq}). Hence, the denominator in
equation~(\ref{eqn:pperi}) involves cancellation between two nearly
equal terms, leading to very large inertial-frame periods.

As Fig.~\ref{fig:period} demonstrates, these long periods are
incompatible with the much-shorter periods ($\sim 1\,\days$) observed
in SPB stars. While the mismatch between theory and observations might
be corrected by considering rotation rates above the present upper
limit $\rfreq/\rfreqc=0.5$, such rapid rotation would be difficult to
reconcile with the angular frequencies $\rfreq/\rfreqc \lesssim 0.25$
more typical to the majority of SPB stars
\citep[e.g.,][]{Aer1999,DeC2005}. The situation is not much improved
for azimuthal orders $m \geq 2$, due to the fact that the instability
regions at $\rfreq/\rfreqc=0.25$ -- a more representative rotation
rate -- are situated at temperatures too cool to encompass many SPB
stars. These findings amount to further evidence against a mixed-mode
interpretation for the SPB phenomenon.

The periods of the unstable $m=1$ mixed modes in Fig.~\ref{fig:period}
are also much longer than those observed in pulsating Be stars. At the
more-rapid rotation rates representative of these stars, it seems
likely that this period discrepancy will be resolved. However, the $m
\geq 2$ mixed modes, which match the typical observed \emph{periods}
quite well when $\rfreq/\rfreqc=0.5$ (even though the \Teff\ range of
the instability is too cool), may then become inconsistent with the
observations. This is one issue that will have to be addressed if
mixed modes are considered a viable explanation for Be-star pulsation.

\section{Discussion and Summary} \label{sec:discuss}

In the preceding sections, I use the non-adiabatic \boojum\ code to
explore the stability of $m=1\ldots4$ mixed modes in rotating B-type
stars. The principal finding is that these modes are excited by the
same iron-bump $\kappa$ mechanism responsible for the pulsation of SPB
and $\beta$ Cep stars. This finding is significant, representing the
discovery of a wholly-new type of instability in the early-type domain
of the HR diagram.

However, I should emphasize certain caveats relating to this
discovery. As discussed in detail in T05, many approximations are
employed to derive the system of pulsation equations and boundary
conditions solved by \boojum. Although these approximations are
required for the problem to be tractable using present-day
computational resources, they do mean that the results presented here
should be regarded as more qualitative than quantitative.

In particular, neglect of the centrifugal deformation of the
equilibrium stellar structure will certainly introduce a degree of
error in the calculated eigenfrequencies. This is because r modes --
and by extension mixed modes -- tend to be rather sensitive to any
departures from a spherically-symmetric equilibrium state
\citep{Pro1981}. Furthermore, the results may be influenced by the
incidence of differential rotation (recently discovered in seismic
analyses of two $\beta$ Cep stars; see \citealp{Dup2004,Pam2004}),
although it is as yet unclear what specific effect any departure from
the assumed-uniform rotation might have.

Nevertheless, while there certainly remains a great deal of scope for
improving upon the theoretical foundation established in the present
work, the case for the physicality of the mixed-mode instability in
B-type stars is, I believe, convincing. Therefore, it is appropriate
to discuss what implications it might have for studies of dynamical
phenomena in early-type stars.

As already made clear in the preceding section, I am skeptical that
mixed modes can furnish a better explanation for the SPB phenomenon
than the currently-accepted model of $\kappa$ mechanism excitation of
(predominantly) $\ell=1$ g modes. The latter model has already proven
very successful both in matching the observed distribution of SPB
stars in the HR diagram \citep[see][]{Pam1999}, and in reproducing the
variability exhibited by these stars
\citep[see][]{Tow2002,DeC2005}. Of course, this is not to say that
there might not be \emph{some} SPB stars that turn out to be mixed
mode rather than g mode pulsators. In particular, the so-called
rapidly rotating Hipparcos SPBs \citep[see][]{Aer1999} may, upon
closer scrutiny, be found to fall into this category. A thorough
evaluation of such a possibility must await new tools for analyzing
the spectroscopic and photometric variations generated by mixed
modes. These tools should prove relatively straightforward to derive
from already-existing models based around the traditional
approximation \citep[see][]{Tow1997,Tow2003b}.

The prospects for a mixed-mode role in pulsating Be stars seem rather
more favourable. Near-critical rotation should shift the instability
strips toward hotter temperatures and earlier spectral types, thereby
allowing them to encompass a greater proportion of these stars. There
may be difficulty in matching periods, and it has yet to be
ascertained whether mixed modes can reproduce observed Be-star line
profile variations to the same degree that \poinc\ modes already have
\citep[see, e.g.,][]{Riv2001,Mai2003}. However, in anticipation of the
eventual resolution of these uncertainties, it is worthwhile
reflecting on the part played by mixed modes in the episodic,
disk-forming mass loss from Be stars \citep[see][and references
therein]{PorRiv2003}. Following on from suggestions made by
\citet{Osa1986}, \citet{Owo2004,Owo2005} has recently advanced a
hydrodynamical model for `pulsationally driven orbital mass ejection'
(PDOME), in which the addition of a pulsation-originated perturbation
to the equatorial regions of a near-critical rotating star is
sufficient to lift surface material into orbit. This material
gradually diffuses outward under the influence of viscosity
\citep[see][and references therein]{Lee1991,Por1999}, to form a disk
with the same Keplerian velocity distribution inferred from
observations of Be stars \citep[e.g.,][]{Dac1986,Han1996}.

As demonstrated by \citet{Owo2005}, a key ingredient in the success of
the PDOME mechanism is that the azimuthal velocity perturbation is in
the prograde direction when the density perturbation is maximal. It
would therefore seem that prograde pulsation modes are required for
mass ejection to occur, which is problematical to reconcile with
observations \citep[][and references therein]{Riv2003} that indicate
retrograde-propagating modes in Be stars. However, an important
distinction should be made here between \emph{group} and \emph{phase}
velocities. While these velocities share the same sign in the case of
\poinc, Kelvin and prograde mixed modes, the retrograde mixed modes
behave differently: their phase velocity is (by definition) in the
retrograde direction, yet their group velocity is oriented in the
opposite direction \citep[see, e.g.,][]{Gil1982}, with maximal density
coinciding with prograde azimuthal velocity. Hence, they are able
simultaneously to fulfill both the observational constraint of
retrograde phase velocity, and the theoretical PDOME constraint of
prograde group velocity. In tandem with a re-evaluation of the degree
to which Be stars approach critical rotation \citep[see][]{Tow2004},
retrograde mixed modes may therefore be the missing link required for
Owocki's PDOME mechanism to operate.

Looking now at the opposite, low-temperature end of the B spectral
type, mixed modes may also have a role to play in the phenomenon of
Maia stars. \citet{Str1955} first put forward the suggestion that
there exists an instability strip in the HR diagram, characterized by
pulsation periods on the order of hours, that extends from late-B
types to the early-A types\footnote{Many authors quote the spectral
range B7-A3, but \citet{Str1955} was never this explicit, resorting
instead to a sketched HR diagram.}. Although the strip was named for
its archetype Maia (20 Tau), subsequent observations
\citep[e.g.,][]{Str1957} of this particular star have, rather
ironically, revealed it most likely to be non-variable. Other searches
for stellar variability in the hypothetical Maia strip
\citep[e.g.,][]{McN1985,Leh1995,PerWil2000} have been largely
unsuccessful. However, there are a handful of variable stars (e.g.,
$\gamma$~CrB -- \citealp{Sch1998}; HD~208727 -- \citealp*{Kal2002})
that remain situated in the strip, and it would therefore be premature
to dismiss the Maia-star concept as mere will-o'-the-wisp.

The discovery of the retrograde mixed-mode instability may provide a
theoretical framework for understanding the origins of the elusive
Maia stars. On the basis of the stellar parameters quoted by
\citet{Kal2002}, the Maia candidate HD~208727 falls blueward of the
ZAMS boundary in the HR diagram (see Fig.~\ref{fig:instab}). However,
if this boundary were shifted to hotter temperatures, for instance by
considering a helium-enriched main sequence, the star would fall well
inside the $m=1\ldots4$ mixed-mode instability regions. Furthermore,
the star's $7.6\,\hours$ photometric period would be consistent with
the inertial frame periods (Fig.~\ref{fig:period}) typical to $m=3$
and $m=4$ mixed modes for $\rfreq/\rfreqc=0.5$ (in fact, the star
probably rotates rather faster than this, $\rfreq/\rfreqc \gtrsim
0.67$). Accordingly, although \citet{Kal2002} were dismissive of the
ability of ordinary g modes to explain the variability of HD~208727,
it appears that a mixed-mode interpretation may furnish a good match.

The interpretation of Maia stars as mixed-mode pulsators is lent
indirect support in a recent paper by \citet{AerKol2005}, who point
out that significant rotation -- a prerequisite for the mixed-mode
instability -- appears to be a common factor in these stars. These
authors then proceed to suggest that the Maia stars may simply be an
extension of the SPB phenomenon to lower \Teff\ arising due to the
Coriolis force. The results presented in T05 would appear to rule out
this latter possibility, by demonstrating (Fig.3, \emph{ibid}) that
any Coriolis-originated extension to the SPB instability strip is
insufficient to encompass the Maia stars. However, as
Figs.~\ref{fig:instab} and~\ref{fig:period} of the present paper
illustrate, the mixed-mode instability at larger azimuthal orders ($m
\gtrsim 4$) \emph{can} encompass cooler temperatures corresponding to
early A spectral types, while at the same time exhibiting pulsation
periods of a few hours. Although this establishes a case for a link
between mixed modes and Maia stars, there should -- as
\citet{AerKol2005} themselves point out -- be a significant
improvement in observational datasets before firm conclusions can be
reached concerning this issue.

Of course, irrespective of whether retrograde mixed modes have any
connection with the Maia phenomenon, there are many pertinent
questions that are raised by the discovery of their $\kappa$-mechanism
instability. What kinds of variability -- spectroscopic or photometric
-- do these modes generate? Although the majority of SPB stars appear
to be well understood as g mode pulsators, are there some that can be
better better explained in terms of mixed mode excitation? Might mixed
modes explain the variability and mass loss of pulsating Be stars? Can
we expect mixed modes in other types of object, such as $\gamma$ Dor
stars? I touch on some of these questions in the foregoing analysis,
and hope further to address them, and others, in forthcoming papers.


\section*{Note added during revision}

A recent preprint by \citet{Sav2005} also investigates mixed-mode
instability in early-type stars. Although there are some differences
in the specific methodology and results, the general findings are in
agreement with those reported in the present work.


\section*{Acknowledgments}

I thank Stan Owocki for useful discussions during the preparation of
the paper, and the anonymous referee for their constructive remarks.
This research has been partially supported by US NSF grant AST-0097983
and NASA grant LTSA04-0000-0060, and by the UK Particle Physics and
Astronomy Research Council.


\bibliography{mixed}

\begin{thebibliography}{}

\bibitem[\protect\citeauthoryear{{Aerts}, {De Cat}, {Peeters}, {Decin}, {De
  Ridder}, {Kolenberg}, {Meeus}, {Van Winckel}, {Cuypers} \&
  {Waelkens}}{{Aerts} et~al.}{1999}]{Aer1999}
{Aerts} C.,  {De Cat} P.,  {Peeters} E.,  {Decin} L.,  {De Ridder} J.,
  {Kolenberg} K.,  {Meeus} G.,  {Van Winckel} H.,  {Cuypers} J.,    {Waelkens}
  C.,  1999, \aap, 343, 872

\bibitem[\protect\citeauthoryear{{Aerts} \& {Kolenberg}}{{Aerts} \&
  {Kolenberg}}{2005}]{AerKol2005}
{Aerts} C.,  {Kolenberg} K.,  2005, \aap, 431, 615

\bibitem[\protect\citeauthoryear{{Andersson}}{{Andersson}}{1998}]{And1998}
{Andersson} N.,  1998, \apj, 502, 708

\bibitem[\protect\citeauthoryear{{Baade}}{{Baade}}{1989a}]{Baa1989a}
{Baade} D.,  1989a, \aaps, 79, 423

\bibitem[\protect\citeauthoryear{{Baade}}{{Baade}}{1989b}]{Baa1989b}
{Baade} D.,  1989b, \aap, 222, 200

\bibitem[\protect\citeauthoryear{{Berthomieu} \& {Provost}}{{Berthomieu} \&
  {Provost}}{1983}]{BerPro1983}
{Berthomieu} G.,  {Provost} J.,  1983, \aap, 122, 199

\bibitem[\protect\citeauthoryear{{Bildsten}, {Ushomirsky} \&
  {Cutler}}{{Bildsten} et~al.}{1996}]{Bil1996}
{Bildsten} L.,  {Ushomirsky} G.,    {Cutler} C.,  1996, \apj, 460, 827

\bibitem[\protect\citeauthoryear{{B\"{o}hm-Vitense}}{{B\"{o}hm-Vitense}}{1981}%
]{Boh1981}
{B\"{o}hm-Vitense} E.,  1981, \araa, 19, 295

\bibitem[\protect\citeauthoryear{{Castor}}{{Castor}}{1971}]{Cas1971}
{Castor} J.~I.,  1971, \apj, 166, 109

\bibitem[\protect\citeauthoryear{{Chauville}, {Zorec}, {Ballereau}, {Morrell},
  {Cidale} \& {Garcia}}{{Chauville} et~al.}{2001}]{Cha2001}
{Chauville} J.,  {Zorec} J.,  {Ballereau} D.,  {Morrell} N.,  {Cidale} L.,
  {Garcia} A.,  2001, \aap, 378, 861

\bibitem[\protect\citeauthoryear{{Cox}, {Morgan}, {Rogers} \& {Iglesias}}{{Cox}
  et~al.}{1992}]{Cox1992}
{Cox} A.~N.,  {Morgan} S.~M.,  {Rogers} F.~J.,    {Iglesias} C.~A.,  1992,
  \apj, 393, 272

\bibitem[\protect\citeauthoryear{{Dachs}, {Hanuschik}, {Kaiser} \&
  {Rohe}}{{Dachs} et~al.}{1986}]{Dac1986}
{Dachs} J.,  {Hanuschik} R.,  {Kaiser} D.,    {Rohe} D.,  1986, \aap, 159, 276

\bibitem[\protect\citeauthoryear{{De Cat} \& {Aerts}}{{De Cat} \&
  {Aerts}}{2002}]{DeC2002}
{De Cat} P.,  {Aerts} C.,  2002, \aap, 393, 965

\bibitem[\protect\citeauthoryear{{De Cat}, {Briquet}, {Daszy{\'
  n}ska-Daszkiewicz}, {Dupret}, {de Ridder}, {Scuflaire} \& {Aerts}}{{De Cat}
  et~al.}{2005}]{DeC2005}
{De Cat} P.,  {Briquet} M.,  {Daszy{\' n}ska-Daszkiewicz} J.,  {Dupret} M.~A.,
  {de Ridder} J.,  {Scuflaire} R.,    {Aerts} C.,  2005, \aap, 432, 1013

\bibitem[\protect\citeauthoryear{{Dupret}, {Thoul}, {Scuflaire}, {Daszy{\'
  n}ska-Daszkiewicz}, {Aerts}, {Bourge}, {Waelkens} \& {Noels}}{{Dupret}
  et~al.}{2004}]{Dup2004}
{Dupret} M.-A.,  {Thoul} A.,  {Scuflaire} R.,  {Daszy{\' n}ska-Daszkiewicz} J.,
   {Aerts} C.,  {Bourge} P.-O.,  {Waelkens} C.,    {Noels} A.,  2004, \aap,
  415, 251

\bibitem[\protect\citeauthoryear{{Dziembowski} \& {Kosovichev}}{{Dziembowski}
  \& {Kosovichev}}{1987}]{DziKos1987}
{Dziembowski} W.,  {Kosovichev} A.,  1987, \acta, 37, 313

\bibitem[\protect\citeauthoryear{{Dziembowski}, {Moskalik} \&
  {Pamyatnykh}}{{Dziembowski} et~al.}{1993}]{Dzi1993}
{Dziembowski} W.~A.,  {Moskalik} P.,    {Pamyatnykh} A.~A.,  1993, \mnras, 265,
  588

\bibitem[\protect\citeauthoryear{{Dziembowski} \& {Pamiatnykh}}{{Dziembowski}
  \& {Pamiatnykh}}{1993}]{DziPam1993}
{Dziembowski} W.~A.,  {Pamiatnykh} A.~A.,  1993, \mnras, 262, 204

\bibitem[\protect\citeauthoryear{{Gautschy} \& {Saio}}{{Gautschy} \&
  {Saio}}{1993}]{GauSai1993}
{Gautschy} A.,  {Saio} H.,  1993, \mnras, 262, 213

\bibitem[\protect\citeauthoryear{{Gill}}{{Gill}}{1982}]{Gil1982}
{Gill} A.~E.,  1982, Atmosphere-Ocean Dynamics.
Academic Press, London

\bibitem[\protect\citeauthoryear{{Hanuschik}}{{Hanuschik}}{1996}]{Han1996}
{Hanuschik} R.~W.,  1996, \aap, 308, 170

\bibitem[\protect\citeauthoryear{{Kallinger}, {Reegen} \& {Weiss}}{{Kallinger}
  et~al.}{2002}]{Kal2002}
{Kallinger} T.,  {Reegen} P.,    {Weiss} W.~W.,  2002, \aap, 388, L37

\bibitem[\protect\citeauthoryear{{Lee}, {Osaki} \& {Saio}}{{Lee}
  et~al.}{1991}]{Lee1991}
{Lee} U.,  {Osaki} Y.,    {Saio} H.,  1991, \mnras, 250, 432

\bibitem[\protect\citeauthoryear{{Lee} \& {Saio}}{{Lee} \&
  {Saio}}{1987}]{LeeSai1987}
{Lee} U.,  {Saio} H.,  1987, \mnras, 224, 513

\bibitem[\protect\citeauthoryear{{Lee} \& {Saio}}{{Lee} \&
  {Saio}}{1997}]{LeeSai1997}
{Lee} U.,  {Saio} H.,  1997, \apj, 491, 839

\bibitem[\protect\citeauthoryear{{Lehmann}, {Scholz}, {Hildebrandt}, {Klose},
  {Panov}, {Reimann}, {Woche} \& {Ziener}}{{Lehmann} et~al.}{1995}]{Leh1995}
{Lehmann} H.,  {Scholz} G.,  {Hildebrandt} G.,  {Klose} S.,  {Panov} K.~P.,
  {Reimann} H.-G.,  {Woche} M.,    {Ziener} R.,  1995, \aap, 300, 783

\bibitem[\protect\citeauthoryear{{Lou}}{{Lou}}{2000}]{Lou2000}
{Lou} Y.,  2000, \apj, 540, 1102

\bibitem[\protect\citeauthoryear{{Maintz}, {Rivinius}, {{\v S}tefl}, {Baade},
  {Wolf} \& {Townsend}}{{Maintz} et~al.}{2003}]{Mai2003}
{Maintz} M.,  {Rivinius} T.,  {{\v S}tefl} S.,  {Baade} D.,  {Wolf} B.,
  {Townsend} R.~H.~D.,  2003, \aap, 411, 181

\bibitem[\protect\citeauthoryear{{Mathias}, {Aerts}, {Briquet}, {De Cat},
  {Cuypers}, {Van Winckel}, {Flanders.} \& {Le Contel}}{{Mathias}
  et~al.}{2001}]{Mat2001}
{Mathias} P.,  {Aerts} C.,  {Briquet} M.,  {De Cat} P.,  {Cuypers} J.,  {Van
  Winckel} H.,  {Flanders.}   {Le Contel} J.~M.,  2001, \aap, 379, 905

\bibitem[\protect\citeauthoryear{{Matsuno}}{{Matsuno}}{1966}]{Mat1966}
{Matsuno} T.,  1966, J. Meteorol. Soc. Japan, 44, 25

\bibitem[\protect\citeauthoryear{{McNamara}}{{McNamara}}{1985}]{McN1985}
{McNamara} B.~J.,  1985, \apj, 289, 213

\bibitem[\protect\citeauthoryear{{Muller}}{{Muller}}{1956}]{Mul1956}
{Muller} D.~E.,  1956, Math. Tab. \& Aids to Comp., 10, 208

\bibitem[\protect\citeauthoryear{{Osaki}}{{Osaki}}{1986}]{Osa1986}
{Osaki} Y.,  1986, \pasp, 98, 30

\bibitem[\protect\citeauthoryear{{Owocki}}{{Owocki}}{2004}]{Owo2004}
{Owocki} S.~P.,  2004, in {Maeder} A.,  {Eenens} P.,  eds, Proc. IAU Symp. 215,
  Stellar Rotation. Astron. Soc. Pac., San Francisco, p.~515

\bibitem[\protect\citeauthoryear{{Owocki}}{{Owocki}}{2005}]{Owo2005}
{Owocki} S.~P.,  2005, in {Ignace} R.,  {Gayley} K.~G.,  eds, ASP Conf. Ser.
  337, The Nature and Evolution of Disks around Hot Stars. Astron. Soc. Pac.,
  San Francisco, in press

\bibitem[\protect\citeauthoryear{{Pamyatnykh}}{{Pamyatnykh}}{1999}]{Pam1999}
{Pamyatnykh} A.~A.,  1999, \acta, 49, 119

\bibitem[\protect\citeauthoryear{{Pamyatnykh}, {Handler} \&
  {Dziembowski}}{{Pamyatnykh} et~al.}{2004}]{Pam2004}
{Pamyatnykh} A.~A.,  {Handler} G.,    {Dziembowski} W.~A.,  2004, \mnras, 350,
  1022

\bibitem[\protect\citeauthoryear{{Papaloizou} \& {Pringle}}{{Papaloizou} \&
  {Pringle}}{1978}]{PapPri1978}
{Papaloizou} J.,  {Pringle} J.~E.,  1978, \mnras, 182, 423

\bibitem[\protect\citeauthoryear{{Percy}, {Harlow} \& {Wu}}{{Percy}
  et~al.}{2004}]{Per2004}
{Percy} J.~R.,  {Harlow} C.~D.~W.,    {Wu} A.~P.~S.,  2004, \pasp, 116, 178

\bibitem[\protect\citeauthoryear{{Percy}, {Hosick}, {Kincaide} \&
  {Pang}}{{Percy} et~al.}{2002}]{Per2002}
{Percy} J.~R.,  {Hosick} J.,  {Kincaide} H.,    {Pang} C.,  2002, \pasp, 114,
  551

\bibitem[\protect\citeauthoryear{{Percy} \& {Wilson}}{{Percy} \&
  {Wilson}}{2000}]{PerWil2000}
{Percy} J.~R.,  {Wilson} J.~B.,  2000, \pasp, 112, 846

\bibitem[\protect\citeauthoryear{{Porter}}{{Porter}}{1996}]{Por1996}
{Porter} J.~M.,  1996, \mnras, 280, L31

\bibitem[\protect\citeauthoryear{{Porter}}{{Porter}}{1999}]{Por1999}
{Porter} J.~M.,  1999, \aap, 348, 512

\bibitem[\protect\citeauthoryear{{Porter} \& {Rivinius}}{{Porter} \&
  {Rivinius}}{2003}]{PorRiv2003}
{Porter} J.~M.,  {Rivinius} T.,  2003, \pasp, 115, 1153

\bibitem[\protect\citeauthoryear{{Press}, {Teukolsky}, {Vetterling} \&
  {Flannery}}{{Press} et~al.}{1992}]{Pre1992}
{Press} W.~H.,  {Teukolsky} S.~A.,  {Vetterling} W.~T.,    {Flannery} B.~P.,
  1992, {Numerical Recipes in Fortran}, 2 edn.
Cambridge University Press, Cambridge

\bibitem[\protect\citeauthoryear{{Provost}, {Berthomieu} \& {Rocca}}{{Provost}
  et~al.}{1981}]{Pro1981}
{Provost} J.,  {Berthomieu} G.,    {Rocca} A.,  1981, \aap, 94, 126

\bibitem[\protect\citeauthoryear{{Rivinius}, {Baade} \& {{\v
  S}tefl}}{{Rivinius} et~al.}{2003}]{Riv2003}
{Rivinius} T.,  {Baade} D.,    {{\v S}tefl} S.,  2003, \aap, 411, 229

\bibitem[\protect\citeauthoryear{{Rivinius}, {Baade}, {{\v S}tefl}, {Townsend},
  {Stahl}, {Wolf} \& {Kaufer}}{{Rivinius} et~al.}{2001}]{Riv2001}
{Rivinius} T.,  {Baade} D.,  {{\v S}tefl} S.,  {Townsend} R.~H.~D.,  {Stahl}
  O.,  {Wolf} B.,    {Kaufer} A.,  2001, \aap, 369, 1058

\bibitem[\protect\citeauthoryear{{Saio}}{{Saio}}{1982}]{Sai1982}
{Saio} H.,  1982, \apj, 256, 717

\bibitem[\protect\citeauthoryear{{Savonije}}{{Savonije}}{2005}]{Sav2005}
{Savonije} G.~J.,  2005, \aap, in press, astro-ph/0506153

\bibitem[\protect\citeauthoryear{{Scholz}, {Lehmann}, {Hildebrandt}, {Panov} \&
  {Iliev}}{{Scholz} et~al.}{1998}]{Sch1998}
{Scholz} G.,  {Lehmann} H.,  {Hildebrandt} G.,  {Panov} K.,    {Iliev} L.,
  1998, \aap, 337, 447

\bibitem[\protect\citeauthoryear{{Stellingwerf}}{{Stellingwerf}}{1979}]{Ste197%
9}
{Stellingwerf} R.~F.,  1979, \apj, 227, 935

\bibitem[\protect\citeauthoryear{{Struve}}{{Struve}}{1955}]{Str1955}
{Struve} O.,  1955, \skytel, 14, 461

\bibitem[\protect\citeauthoryear{{Struve}, {Sahade}, {Lynds} \&
  {Huang}}{{Struve} et~al.}{1957}]{Str1957}
{Struve} O.,  {Sahade} J.,  {Lynds} C.~R.,    {Huang} S.~S.,  1957, \apj, 125,
  115

\bibitem[\protect\citeauthoryear{{Townsend}}{{Townsend}}{1997}]{Tow1997}
{Townsend} R.~H.~D.,  1997, \mnras, 284, 839

\bibitem[\protect\citeauthoryear{{Townsend}}{{Townsend}}{2002}]{Tow2002}
{Townsend} R.~H.~D.,  2002, \mnras, 330, 855

\bibitem[\protect\citeauthoryear{{Townsend}}{{Townsend}}{2003a}]{Tow2003a}
{Townsend} R.~H.~D.,  2003a, \mnras, 340, 1020

\bibitem[\protect\citeauthoryear{{Townsend}}{{Townsend}}{2003b}]{Tow2003b}
{Townsend} R.~H.~D.,  2003b, \mnras, 343, 125

\bibitem[\protect\citeauthoryear{{Townsend}}{{Townsend}}{2005}]{Tow2005}
{Townsend} R.~H.~D.,  2005, \mnras, 360, 465 (T05)

\bibitem[\protect\citeauthoryear{{Townsend}, {Owocki} \& {Howarth}}{{Townsend}
  et~al.}{2004}]{Tow2004}
{Townsend} R.~H.~D.,  {Owocki} S.~P.,    {Howarth} I.~D.,  2004, \mnras, 350,
  189

\bibitem[\protect\citeauthoryear{{Unno}, {Osaki}, {Ando}, {Saio} \&
  {Shibahashi}}{{Unno} et~al.}{1989}]{Unn1989}
{Unno} W.,  {Osaki} Y.,  {Ando} H.,  {Saio} H.,    {Shibahashi} H.,  1989,
  {Nonradial Oscillations of Stars}, 2 edn.
University of Tokyo Press, Tokyo

\bibitem[\protect\citeauthoryear{{Waelkens}, {Aerts}, {Kestens}, {Grenon} \&
  {Eyer}}{{Waelkens} et~al.}{1998}]{Wae1998}
{Waelkens} C.,  {Aerts} C.,  {Kestens} E.,  {Grenon} M.,    {Eyer} L.,  1998,
  \aap, 330, 215

\bibitem[\protect\citeauthoryear{{Yanai} \& {Maruyama}}{{Yanai} \&
  {Maruyama}}{1966}]{YanMur1966}
{Yanai} M.,  {Maruyama} T.,  1966, J. Meteorol. Soc. Japan, 44, 291

\end{thebibliography}

\bibliographystyle{mn2e}


\label{lastpage}

\end{document}